\documentclass[aps,amsfont,amssymb,twocolumn,floats,prl,superscriptaddress,floatfix]{revtex4}
\usepackage{epsfig,graphicx}
\newcommand{\lnG}{\widetilde{\log}}
\newcommand{\expG}{\widetilde{\exp}}
\newcommand{\lambdaG}{\lambda}
\newcommand{\alphasens}{\alpha^{\mathrm{av}}_{\mathrm{sens}}}
\begin{document}
%
\title{
Entropy production and Pesin-like identity at the onset of chaos}
\author{Roberto Tonelli}
\email{roberto.tonelli@dsf.unica.it}
\affiliation{Dipart. di Fisica dell'Universit\`a di Cagliari,
             I-09042 Monserrato, Italy}
\affiliation{INFO-SLACS Laboratory, I-09042 Monserrato, Italy} 
\affiliation{Ist. Naz. Fisica Nucleare (I.N.F.N.) Cagliari,
             I-09042 Monserrato, Italy}
\author{Giuseppe Mezzorani}
\email{giuseppe.mezzorani@ca.infn.it}
\affiliation{Dipart. di Fisica dell'Universit\`a di Cagliari,
             I-09042 Monserrato, Italy}
\affiliation{Ist. Naz. Fisica Nucleare (I.N.F.N.) Cagliari,
             I-09042 Monserrato, Italy}
\author{Franco Meloni}
\email{franco.meloni@dsf.unica.it} \affiliation{Dipart. di Fisica
dell'Universit\`a di Cagliari,
             I-09042 Monserrato, Italy}
\affiliation{INFO-SLACS Laboratory, I-09042 Monserrato, Italy}
\author{Marcello Lissia}
\email{marcello.lissia@ca.infn.it}
\affiliation{Ist. Naz. Fisica Nucleare (I.N.F.N.) Cagliari,
             I-09042 Monserrato, Italy}
\affiliation{Dipart. di Fisica dell'Universit\`a di Cagliari,
             I-09042 Monserrato, Italy}
\author{Massimo Coraddu}
\email{massimo.coraddu@ca.infn.it}
\affiliation{Dipart. di Fisica dell'Universit\`a di Cagliari,
             I-09042 Monserrato, Italy}
\affiliation{Ist. Naz. Fisica Nucleare (I.N.F.N.) Cagliari,
             I-09042 Monserrato, Italy}
\date{December 28, 2004; Last revision November 3, 2005}
%
\begin{abstract}
Asymptotically entropy of chaotic systems increases linearly and
the sensitivity to initial conditions is exponential with time:
these two behaviors are related. Such relationship is the
analogous of and under specific conditions has been shown to
coincide with the Pesin identity. Numerical evidences support the
proposal that the statistical formalism can be extended to the
edge of chaos by using a specific generalization of the
exponential and of the Boltzmann-Gibbs entropy. We extend this
picture and a Pesin-like identity to a wide class of deformed
entropies and exponentials using the logistic map as a test case.
The physical criterion of finite-entropy growth strongly restricts
the suitable entropies. The nature and characteristics of this
generalization are clarified.
\end{abstract}
 \maketitle
It has been suggested that exponential and power-law sensitivity
to initial conditions of chaotic systems could be unified through
the concept of generalized exponential leading to the definition
of generalized Lyapunov exponents~\cite{Tsallis:1997}: the
sensitivity $ \xi\equiv \lim_{\Delta x(0)\to 0} \Delta x(t) /
\Delta x(0)$ grows asymptotically as a generalized exponential $
\expG(\lambdaG t)$, where $\expG(x) = \exp_q(x) \equiv
[1+(1-q)x]^{1/(1-q)}$; the exponential behavior for the chaotic
regime is recovered for $q\to1$: $\lim_{q\to1} \exp_q(\lambda_q t)
= \exp(\lambda t)$.

The Kolmogorov-Sinai (K-S) entropy is defined as the time-averaged
production rate of Shannon entropy over the joint probability for
the trajectory to pass through a given sequence of steps in phase
space with appropriate coding of the sequence and limit of the
volume that characterize the steps. The K-S definition for most
systems is not easily amenable to numerical evaluation and not
directly equivalent to the usual thermal or statistical entropy.
The connection between the K-S entropy production rate and the
statistical entropy of fully chaotic systems and, in particular,
of the three successive stages in the entropy time evolution is
discussed in Ref.~\cite{Latora:1999prl}. After a first
initial-condition-dependent stage, the entropy grows linearly in
the limit of infinitely fine coarse graining: this is the
characteristic asymptotic regime. The last stage depends on the
practical limitation of the coarse graining.

Since this statistical definition of entropy production rate,
where an ensemble of initial conditions confined to a small region
is let evolve, appears to be relevant to thermal processes and
practically coincides with the K-S entropy in chaotic
regimes~\cite{Latora:1999prl}, we use it and call from now one K-S
entropy $K$, even if it is in principle different.

The rate of loss of information relevant for the above-mentioned
systems, which include both full-chaotic and edge-of-chaos cases,
should be suitably generalized. In fact at the edge of chaos this
picture appears to be recovered with the generalized entropic form
proposed by Tsallis~\cite{Tsallis:1987eu} $S_q = (1-\sum_{i=1}^{W}
p_i^q)/(q-1)$, which grows linearly for a specific value of the
entropic parameter $q=q_{\mathrm{sens}}=0.2445$ for the logistic
map:
 $\lim_{t\to\infty}\lim_{L\to 0} S_q(t)/ t = K_q$, where $S_q $
reduces to $-\sum_{i=1}^{W} p_i\log p_i$ in the limit $q\to 1$
$p_i$ being the fraction of the ensemble found in the $i$-th cell
of linear size $L$.
 The same
exponent describes the asymptotic power-law sensitivity to initial
conditions \cite{Latora:1999vk}.

Finally, it has been also conjectured that the relationship
between the asymptotic entropy-production rate $K$ and the
Lyapunov exponent $\lambda$ for  chaotic systems (Pesin-like
identity) can be extended to $K_q =
\lambda_q$~\cite{Tsallis:1997}.

 Numerical evidences supporting
this framework with the entropic form
 $S_q $ have been found for the
logistic~\cite{Tsallis:1997} and generalized logistic-like
maps~\cite{Tsallis:1997cl}. Ref.~\cite{Lyra:1998wz} gives scaling
arguments that relate the asymptotic power-law behavior of the
sensitivity to the fractal scaling properties of the attractor.
Finite-size scaling relates numerically the entropic parameter $q$
to the power-law convergence to the box measure of the attractor
from initial conditions uniformly spread over phase space
\cite{deMoura:2000,Borges:2002}. The linear behavior of the
generalized entropy  for specific choices of $q$ has been observed
for two families of one-dimensional dissipative maps
\cite{Tirnakli:2001}.

Renormalization-group methods have been recently
used~\cite{Baldovin:2002a,Baldovin:2002b} to demonstrate that the
asymptotic (large $t$) behavior of the sensitivity to initial
conditions in the logistic and generalized logistic maps is
bounded by a specific power-law whose exponent could be determined
analytically for initial conditions on the attractor. Again for
evolution on the attractor the connection between entropy and
sensitivity has been studied analytically and numerically,
validating a Pesin-like identity for Tsallis' entropic form\
\cite{Baldovin:2004}.

Sensitivity and entropy production have been studied in
one-dimensional dissipative maps using ensemble-averaged initial
conditions chosen uniformly over the interval embedding the
attractor~\cite{Ananos:2004a}: the statistical picture, i.e., the
relation between sensitivity and entropy exponents and the
Pesin-like identity, has been confirmed even if with a different
$q=q_{\mathrm{sens}}^{\mathrm{ave}}=0.35$~\cite{Ananos:2004a}.
Indeed the ensemble-averaged initial conditions appear more
relevant for the connection between ergodicity and chaos and for
practical experimental settings. The same analysis has been
repeated for two symplectic standard maps~\cite{Ananos:2004b}.

The main objective of the present work is to demonstrate the more 
general validity of the above-described picture and, specifically, 
of the extended Pesin-like identity and of the physical criterion 
of finite-entropy production per unit time (linear growth), which 
strongly selects appropriate entropies and fixes their parameters. 
For this purpose we use the coherent statistical framework arising 
from the following two-parameter 
family~\cite{Mittal,Taneja1,Borges1,Kaniadakis:2004nx,Kaniadakis:2004ri,Kaniadakis:2004td} 
of logarithms
\begin{equation}\label{eq:logGen}
    \lnG(\xi) \equiv \frac{\xi^{\alpha}-\xi^{-\beta}}{\alpha+\beta}
    \quad ,
\end{equation}
where $\alpha$ ($\beta$) characterizes the large (small) argument
asymptotic behavior;  in particular, $\expG(\lambdaG t)\equiv
\lnG^{-1}(\lambdaG t)\sim (\lambdaG t)^{1/\alpha}$ for large $t$.
The requirements that the logarithm be an increasing function,
therefore invertible, with negative concavity (the related entropy
is stable), and that the exponential be normalizable for negative
arguments (see Ref.~\cite{Naudts1} for a discussion of the
criteria) select $0 \leq \alpha\leq 1 $ and $0\leq \beta <1
$~\cite{Kaniadakis:2004td}; solutions with negative $\alpha$ and
$\beta$ are duplicate, since the family (\ref{eq:logGen}) 
possesses the symmetry $\alpha \leftrightarrow -\beta $. We have 
studied the entire class of logarithms and exponentials and will 
show results for the following four interesting one-parameter 
cases:

(1) the original Tsallis proposal~\cite{Tsallis:1987eu}:
$\alpha=1-q$ and $\beta=0$;

(2) the Abe logarithm: $q_A=1/(1+\alpha)$ and
$\beta=\alpha/(1+\alpha)$, named after the entropy introduced by
Abe~\cite{Abe:1997qg}, which possesses the symmetry $q_A\to
1/q_A$;

(3) the Kaniadakis logarithm: $\alpha=\beta=\kappa$, which shares
the same symmetry group of the relativistic momentum
transformation and has applications in cosmic-ray and plasma
physics~\cite{Kaniadakis:2001nl,Kaniadakis:2002sr,Kaniadakis:2005zk};

(4) in addition we have considered a sample spanning the entire
class (values of $\beta$): in this letter we show results for the
choice $\alpha=2\beta=2\gamma$, which we label $\gamma$
logarithm~\footnote{The exponentials corresponding to the
$\gamma$, Tsallis, and Kaniadakis logarithms,
$\expG(y)=\lnG^{-1}(y)$, are among the members of the class that
can be explicitly inverted in terms of simple known
functions~\cite{Kaniadakis:2004td}.}.

We study the sensitivity to initial conditions and the entropy
production in the logistic map $x_{i+1}=1-\mu x^2_i$, at the
infinite bifurcation point $\mu_{\infty}=1.401155189$. If the
sensitivity  follows a deformed exponential $\xi(t) =
\expG(\lambdaG t)$, analogously to the chaotic regime when
$\xi(t)\sim \exp(\lambda t)$, the corresponding deformed logarithm
of $\xi$  should yield a straight line
$\lnG(\xi(t))=\lnG(\expG(\lambdaG t))=\lambdaG t$.

The sensitivity has been calculated using the exact formula for
the logistic map $\xi(t)= (2\mu)^{t}\prod_{i=0}^{t-1}|x_{i}|$ for
$1 \leq t \leq 80 $, and the generalized logarithm $\lnG(\xi)$ has
been uniformly averaged over the entire interval $-1<x_0<1$ by
randomly choosing $4\times 10^{7}$ initial conditions. The
averaging over initial conditions, indicated by
$\langle\cdots\rangle$, is appropriate for a comparison with the
entropy production.

Guided by Ref.~\cite{Ananos:2004a}, for each of several
generalized logarithms, $\langle\lnG(\xi(t))\rangle$ has been
fitted to a quadratic function for $1\leq t \leq 80$ and $\alpha$
has been chosen such that the coefficient of the quadratic term be
zero; in fact $\lnG(\xi)$ linear in $t$ means that the sensitivity
$\xi$ behaves as $\expG(\lambda t)$: we label this value
$\alphasens$.

In fact, the exponent obtained with this procedure has been
denoted $q_{\mathrm{sens}}^{\mathrm{av}}$ in the case of Tsallis'
entropy~\cite{Ananos:2004a} to mark the difference with the
exponent $q_{\mathrm{sens}}$ obtained by choosing an initial
condition at the fixed point of the map
$x_0=0$~\cite{Baldovin:2004}.

Table~\ref{tab:pesin} reports the values of $\alphasens$
corresponding to four different choices of the logarithm with the
shown statistical errors calculated by repeating the fitting
procedure for sub-samples of the $4\times 10^{7}$ initial
conditions; in addition we estimate a systematic error of about
$0.004$ by fitting over different ranges of $t$. The exponent we
find using Tsallis' formulation $\alphasens = 0.644\pm 0.002$ is
consistent within the errors with the value quoted in
Ref.~\cite{Ananos:2004a}
$q^{\mathrm{av}}_{\mathrm{sens}}=1-\alphasens\approx 0.36$. The
values of $\alphasens$ are within $\pm1\%$: taking into account
the systematic error arising from the inclusion in the global
fitting of small values of $\xi$, for which the different
formulations have not yet reached their common asymptotic behavior
(at least at the level of 1\%), it can be estimated
$\alphasens=0.650\pm 0.005$.

\begin{table*}[htb]
\caption{For each of the generalized logarithms listed in the
first row the following quantities are given: the exponents of the
leading-power behavior for large- ($\alpha$) and small- ($\beta$)
arguments according to Eq.~(\ref{eq:logGen}); the experimental
generalized Lyapunov exponents $\lambda$ (see
Fig.~\ref{fig:sensitivity}) and the experimental rates of entropy
growth $K$ (see Fig.~\ref{fig:entropy}), which verify the
generalized Pesin-like identity $K=\lambda$; the asymptotic
($\nu_{\mathrm{eff}}=\infty$) and corrected
($\nu_{\mathrm{eff}}=20$) theoretical predictions for $K_{\beta}$
relative to $K_0$ ($\beta=0$ or Tsallis' case) from
Eq.~(\ref{eq:KBeta}), and the corresponding experimental values.
Statistical errors are shown. \label{tab:pesin}}
\newcommand{\dg}{\hphantom{$0$}}
\newcommand{\cc}[1]{\multicolumn{1}{c}{#1}}
\renewcommand{\tabcolsep}{1pc} 
\begin{tabular}{lcccc}
\hline   & Tsallis& $\gamma$ & Abe & Kaniadakis \\
\hline
 $\alpha$ & $0.644\pm 0.002\ $ & $0.656\pm 0.002 $ & $0.657\pm
0.002$& $0.653\pm 0.002$\\
$\beta$ & 0 & 0.328& 0.396 & 0.653 \\
\hline \hline $\lambda$
&$\mathbf{0.271}\pm0.003$ & $\mathbf{0.198}\pm0.002$ & $\mathbf{0.185}\pm0.002$ &$\mathbf{0.148}\pm0.001$ \\
K &$\mathbf{0.267}\pm0.003$ &$\mathbf{0.197}\pm0.002$  &
$\mathbf{0.186}\pm0.002$& $\mathbf{0.152}\pm0.002$\\ \hline\hline
$\left(K/K_0\right)_{\mathrm{Asymp}}$  & 1 & 0.667& 0.623 & 0.500  \\
\hline
$\left(K/K_0\right)_{\mathrm{Theory}}$ & 1 &0.736 &0.695 & 0.571 \\
$\left(K/K_0\right)_{\mathrm{Experim}}$  & 1 & $0.738\pm0.004$& $0.697\pm 0.004$& $0.569\pm 0.004$\\
\hline
\end{tabular}\\[2pt]
\end{table*}

\begin{figure}[ht]
\includegraphics[width=\columnwidth,angle=0]{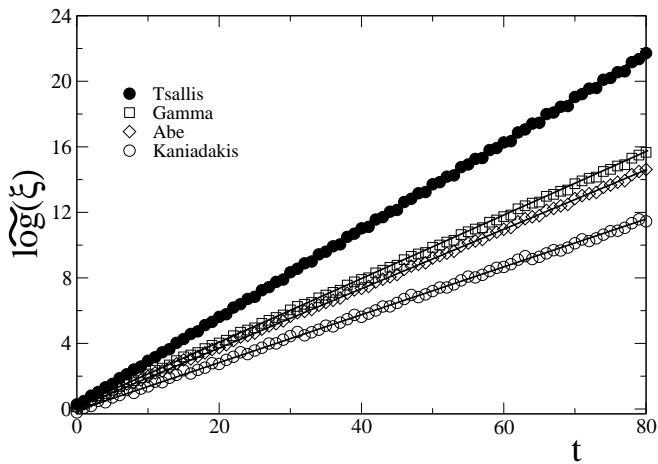}
\caption{Generalized logarithms $\lnG(\xi)$, see
Eq.~(\ref{eq:logGen}), of the sensitivity to initial conditions
averaged over $4\times 10^{7}$ uniformly distributed initial
conditions as function of time. From top to bottom: Tsallis',
$\gamma$, Abe's, and Kaniadakis' logarithms. The linear rising
behavior has been obtained with the asymptotic powers $\alpha$
shown in Table~\ref{tab:pesin}; the corresponding generalized
Lyapunov exponents $\lambda$ can also be read in
Table~\ref{tab:pesin}. \label{fig:sensitivity}}
\end{figure}

Figure~\ref{fig:sensitivity} shows the straight-line behavior of
$\lnG(\xi)$  for all four formulations when $\alpha=\alphasens$:
 the corresponding slopes $\lambda$ (generalized Lyapunov
exponents) are reported in Table~\ref{tab:pesin} with their
statistical errors; a systematic error of about 0.003 has been
estimated by different choices of the range of $t$. While the
values of $\alpha$ are consistent with a universal exponent
independent of the particular deformation, the slope $\lambda$
strongly depends on the choice of the logarithm.

The entropy has been calculated by dividing the interval $(-1,1)$
in $W=10^{5}$ equal-size boxes, putting at the initial time
$N=10^{6}$ copies of the system with an uniform random
distribution within one box, and then letting the systems evolve
according to the map. At each time $p_i(t)\equiv n_i(t)/N$, where
$n_i(t)$ is the number of systems found in the $i$-th box at time
$t$, the entropy of the ensemble is
\begin{equation}\label{eq:entropyGen}
    S(t) \equiv \left\langle \sum_{i=1}^{W} p_i(t) \lnG(\frac{1}{p_i(t)})\right\rangle =
    \left\langle \sum_{i=1}^{W} \frac{p_i^{1-\alpha}(t)-p_i^{1+\beta}(t)}{\alpha+\beta}
         \right\rangle
\end{equation}
where $\langle\cdots\rangle$ is an average over $2\times 10^{4}$
experiments, each one starting from one box randomly chosen among
the $W$ boxes. The choice of the entropic
form~(\ref{eq:entropyGen}) is fundamental for a coherent
statistical picture: the usual constrained variation of the
entropy in Eq.~(\ref{eq:entropyGen}) respect to $p_i$ yields as
distribution the deformed exponential $\expG(x)$ whose inverse is
indeed the logarithm appearing in
Eq.~(\ref{eq:logGen})~\cite{Kaniadakis:2004td}.

\begin{figure}[ht]
\includegraphics[width=0.8\columnwidth,angle=-90]{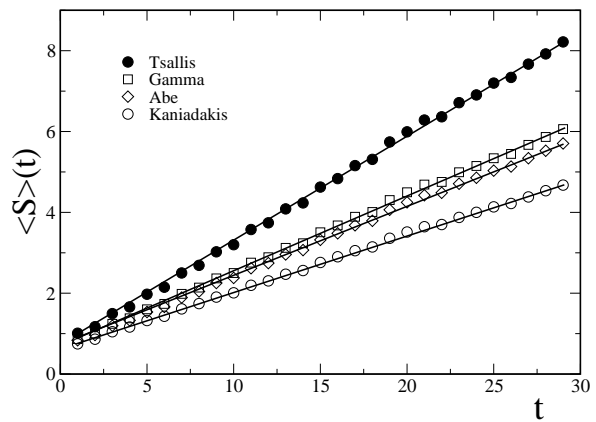}
\caption{Entropy as function of time, averaged over $2\times
10^{4}$ experiments, each one with an ensemble of $10^{6}$ points
over a grid of $10^5$ boxes. The entropies belong to the class of
Eq.~(\ref{eq:entropyGen}) and are defined with exactly the same
exponents $\alpha$ and $\beta$ used for the sensitivities; as in
Fig.~\ref{fig:sensitivity},  the curves show Tsallis', $\gamma$,
Abe's, and  Kaniadakis' entropies from top to bottom. Straight
lines are guides to the eyes. The resulting rates of growth of
entropy $K$ can be read in Table~\ref{tab:pesin}.
\label{fig:entropy} }
\end{figure}

Analogously to the strong chaotic case, where an exponential
sensitivity ($\alpha=\beta=0$) is associated to a linear rising
Shannon entropy, which is defined in terms of the usual  logarithm
($\alpha=\beta=0$), and consistently with the conjecture in
Ref.~\cite{Tsallis:1997}, the same values $\alpha$ and $\beta$ of
the sensitivity are used in Eq.~(\ref{eq:entropyGen}):
Fig.~\ref{fig:entropy} shows that this choice leads to entropies
that grow also linearly. We have also verified that this linear
behavior is lost for values of the exponent $\alpha$ different
from $\alphasens$, confirming for the whole
class~(\ref{eq:entropyGen}) what was already known for the
$q$-logarithm~\cite{Tsallis:1997,Ananos:2004a}.

In Table~\ref{tab:pesin} we report the rate of growth of several
entropies $K_{\beta}=S_{\beta}(t)/t$; the statistical errors have
been again estimated by sub-sampling the experiments. While this
rate $K$ depends on the choice of the entropy, the Pesin-like
identity holds for each given deformation:
$K_{\beta}=\lambda_{\beta}$.

There exists an intuitive explanation of the dependence of the
value of $K$ on $\beta$, i.e., on the choice of the deformation.
If the ensemble at time $t$ is spread approximately uniformly over
$\nu$ boxes, while the other $(W-\nu)$ boxes are practically empty
the entropy is $  S(t)\approx \lnG(\nu(t)) \sim
    [\nu^{\alpha}(t)-\nu^{-\beta}(t)]/(\alpha+\beta)$, being
    $\nu$ an effective average number of occupied boxes.
Since the exponent $\alpha$ is a property of the map, and
practically independent of the entropic form, and choosing the
rate of growth of Tsallis' entropy $K_0$, which  corresponds to
$\beta=0$, as reference,
 the dependence of $K_{\beta}$ on the choice of $\beta$ results:
\begin{equation}\label{eq:KBeta}
   \frac{K_{\beta}}{K_0}
    = \left(1 +\frac{\beta}{\alpha}\right)^{-1} \times
    \frac{1-\nu^{-\alpha-\beta}}{1-\nu^{-\alpha}}\quad .
\end{equation}
The first factor is the asymptotic $(\nu\gg 1)$ ratio that depends
only on $\alpha$ and $\beta$, while the second factor gives the
leading finite size correction. The last three lines of
Table~\ref{tab:pesin} compare the asymptotic and corrected
theoretical predictions of Eq.~(\ref{eq:KBeta}) with the numerical
experiment: this intuitive picture appears to hold with a 10\%
discrepancy between the asymptotic value and the corrected one
which in turns reproduces accurately the experimental results with
an effective number of occupied boxes $\nu \approx 20$. This
result is clearly not restricted to the four representative cases
explicitly shown.

In summary, numerical evidence corroborates and extends Tsallis' 
conjecture that, analogously to strongly chaotic systems,  also 
weak chaotic systems can be described by an appropriate 
statistical formalism. Such extended formalisms should verify 
precise requirements (concavity, Lesche 
stability~\cite{Scarfone:2004ls}, and finite-entropy production 
per unit time) to both correctly describe chaotic systems and 
provide a coherent statistical framework: especially the last 
criterion restricts the  entropic forms to the ones with the 
correct asymptotic behavior. These results have been exemplified 
in a specific two-parameter class that meets all these 
requirements: it is the simplest power-law form describing small 
and large probability behavior and includes Tsallis's seminal 
proposal. More specifically, the logistic map shows

(i) a power-low sensitivity to initial condition with a specific
exponent $\xi\sim t^{1/\alpha}$, where $\alpha = 0.650\pm 0.005$;
this sensitivity  can be described by deformed exponentials with
the same asymptotic behavior $\xi(t)=\expG(\lambdaG t)$ (see
Fig.~\ref{fig:sensitivity} for examples);

(ii) a constant asymptotic entropy production rate (see
Fig.~\ref{fig:entropy}) for trace-form entropies with a specific
power-behavior $p^{1-\alpha}$ in the limit of small probabilities
only when the exponent $\alpha$ is the same appearing in the
sensitivity;

(iii) the asymptotic exponent $\alpha$ is related to parameters of
known entropies: for instance $\alpha = 1-q$, where $q$ is the
entropic index of Tsallis' thermodynamics~\cite{Tsallis:1987eu};
$\alpha=1/q_A-1$, where $q_A$ appears in a generalization of Abe's
entropy~\cite{Abe:1997qg}; $\alpha =\kappa$, where $\kappa$ is the
parameter in Kaniadakis'
statistics~\cite{Kaniadakis:2001nl,Kaniadakis:2002sr,Kaniadakis:2005zk};

(iv) a generalized Pesin-like identity holds $S_{\beta}/t\to
K_{\beta} = \lambda_{\beta}$ for each choice of entropy and
corresponding exponential in the class, even if the value of
$K_{\beta}=\lambda_{\beta}$ depends on the specific entropy (or
choice of $\beta$) and it is not characteristic of the map as it
is $\alpha$ (see Table~\ref{tab:pesin} for examples);

(v) the ratios between  the entropy production rates $K_{\beta}$
(analogous of the K-S entropy) from different members of the class
of entropies in Eq.~(\ref{eq:entropyGen}) can be theoretically
understood from the knowledge of the power behavior of the
deformed logarithm for large and small values of the argument,
Eq.~(\ref{eq:KBeta}), and the predictions are confirmed by
numerical experiments (see Table~\ref{tab:pesin}).

Weakly chaotic systems can be characterized by deformations of
statistical mechanics that yield entropic forms with the
appropriate power-law asymptotic ($p_i\to 0$) behavior and a
consequent asymptotic power-law behavior of the corresponding
exponential.

We remark that the physical criterion of requiring that the
entropy production rate reaches a finite and non-zero asymptotic
value has two consequences: (a) it  selects a specific value of
the parameter $\alpha$ ($\alpha$ is characteristic of the system);
(b) strongly restricts the kind of acceptable entropies to the
ones that have asymptotic power-law behavior (for instance it
excludes Renyi entropy~\cite{Johal:2004,Lissia:2005by}). The
reason we ask for a finite non-zero slope is that otherwise we
would miss an important characteristic of the system: its
asymptotic exponent.

The proper generalization of the Pesin-like identity involves two
distinct points: (1) the correspondence between the power-law
behavior of the sensitivity for large times and of the proper
entropy for small occupation probability (this power-law is
independent of the generalization chosen); and (2) the equality of
the generalized Lyapunov exponent and the entropy production rate:
the numerical value is in this case dependent on the choice of the
entropy.

\begin{acknowledgments}
We acknowledge useful comments from S. Abe, F. Baldovin, G.
 Kaniadakis, A. M. Scarfone, U.~Tirnakly, and C. Tsallis.
The present version benefitted by comments of the referee.

This work was partially supported by MIUR (Ministero
dell'Istruzione, dell'Universit\`a e della Ricerca) under
MIUR-PRIN-2003 project ``Theoretical Physics of the Nucleus and
the Many-Body Systems.''
\end{acknowledgments}

\end{document}